\documentclass[a4paper,12pt]{article}
 \usepackage{subcaption}
    \usepackage{graphicx}
    \usepackage[utf8]{inputenc}\usepackage{graphicx}
\usepackage{graphics}
\usepackage[margin=1in]{geometry}
\usepackage{amsthm}
\usepackage{amssymb}
\usepackage{amsmath}
\usepackage[table]{xcolor}
\usepackage{setspace}
\usepackage{url}
\usepackage{mathrsfs}  
\usepackage[figuresright]{rotating}
\usepackage[utf8]{inputenc}
\usepackage{mathtools}
\usepackage{booktabs}
\usepackage{multirow}
\usepackage{chngpage}
\usepackage{bbm}
\usepackage{amsbsy}
\usepackage[round]{natbib}

\newcommand{\RN}[1]{%
  \textup{\uppercase\expandafter{\romannumeral#1}}%
}

\newcommand{\Ccal}{\mathcal{C}}
\newcommand{\Lcal}{\mathcal{L}}
\newcommand{\ncol}{n_{\Ccal } }

\usepackage{manyfoot}

\SelectFootnoteRule{A}

\DeclareNewFootnote{A}

\makeatletter
\def\and{%
  \end{tabular}%
  \begin{tabular}[t]{c}}%
\def\@fnsymbol#1{\ensuremath{\ifcase#1\or 1\or 2\or c\or
   d\or e\or f\or g\or h\or i\else\@ctrerr\fi}}
\makeatother

\title{
A spatio-temporal model and inference tools for longitudinal count data on multicolor cell growth}
\author{Puxue Qiao,\thanks{School of Mathematics and Statistics,  University of Melbourne, VIC,
3010, Australia} \and Christina M{\o}lck, \thanks{Department of Pathology,  University of Melbourne,  VIC, 3010,
Australia} \and
     Davide Ferrari, \footnotemark[1] \and
  Fr\'{e}d\'{e}ric Hollande \footnotemark[2]
   } 
   \date{\vspace{-5ex}}

\begin{document}
\maketitle 
\let\thefootnoteA\relax\footnoteA{\hspace{-0.1 cm}Davide Ferrari,School of Mathematics and Statistics,  University of Melbourne, VIC,
3010, Australia. }
\let\thefootnoteA\relax\footnoteA{\hspace{-0.2 cm} Email: dferrari@unimelb.edu.au }

\begin{abstract}
Multicolor cell spatio-temporal image data have become important to investigate
organ development and regeneration, malignant growth or immune responses by tracking
different cell types both in vivo and in vitro. 
Statistical modeling of image data from common longitudinal cell experiments
poses significant challenges due to the presence of complex spatio-temporal
interactions between different cell types and difficulties related to measurement of single cell trajectories. 
Current analysis methods focus mainly on univariate cases, often not considering the spatio-temporal effects 
affecting cell growth between different cell populations.  
In this paper, we propose a conditional spatial autoregressive model
to describe multivariate count cell data on the lattice, and develop inference tools. The proposed methodology is
computationally tractable and enables researchers to estimate a complete statistical model of multicolor cell growth. 
Our methodology is applied on real experimental data where we investigate how interactions between cells affect their growth. We include two case studies; the first evaluates interactions between cancer cells and fibroblasts, which are normally present in the tumor microenvironment, whilst the second evaluates interactions between cloned cancer cells when grown as different combinations.
\end{abstract}

Keywords:
Spatio-temporal lattice model, count data, multicolor cell growth 

\section{Introduction}
\label{s:intro}
Longitudinal image data based on fluorescent proteins play a crucial role for
both in vivo and in vitro analysis of various biological processes 
such as gene expression and cell lineage fate. Assessing the growth patterns of
different cell types within a heterogeneous population and monitoring their 
 interactions enables biomedical researchers to determine the role of different cell types
 in important biological processes such as  organ development and regeneration, malignant growth or 
immune responses under various experimental conditions. 
For example, tumor progression has been shown to be affected by bidirectional interactions among cancer cells or between cancer cells and cells from the microenvironment, including tumor-infiltrating immune cells
 \cite{medema2011microenvironmental}. Being able to study these interactions in a laboratory setting is therefore highly relevant, but is complicated by the difficulty of dissecting the effect of the different cell types as soon as the number of cell types exceeds two. In the present study we used longitudinal image data collected from multicolor live-cell imaging growth experiments of co-cultures of cancer cells and fibroblasts (a key cell type in the tumor microenvironment) as well as behaviourally distinct (cloned) cancer cells. Using a high-content imaging system, we were able to acquire characteristics for each individual cell at subsequent times, including fluorescent properties, spatial coordinates, and morphological features. The motivation of this work was to design a model allowing the determination of spatio-temporal growth interactions between these multiple cell populations.
 
In longitudinal growth experiments, the two important goals are to determine growth
rates for different cell populations and to assess how 
interactions between cell types may affect their growth. Whilst a wide range of
descriptive data analysis approaches have been used in applications, inference based on a comprehensive model
of multicolor cell data is an open research area. The main challenges are related to the
presence of complicated spatio-temporal interactions amongst cells and
difficulties related to tracking individual
cells across time from image data. Typical longitudinal
experiments consist of a relatively small number of measurements (e.g. 5 to 20 images taken every few hours),
which is adequate for monitoring cell growth. 
Tracking individual cells would typically require more frequent measurements, complicating the practicality of the experiments in terms of the storage cost of very large image files and the cytotoxicity induced by the imaging process.

\begin{figure}
\includegraphics[scale = 0.9]{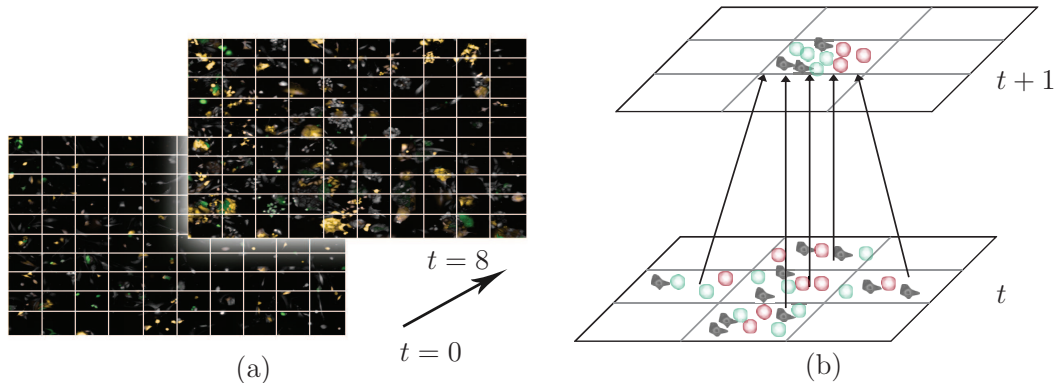}
\caption{(a) Microscope images for the cancer cell growth data  obtained from a
high-content imager (Operetta, 
Perkin Elmer) at the initial and final time points of the experiment. In each
image, colors for non-fluorescent fibroblasts, as well as red and green fluorescent
cancer cells are merged. (b) Illustration of the local structure for the model in (\ref{intensity1}). The two planes correspond to $3 \times
3$ tiles at times 
$t$ and $t+1$. The average number of cells of  color  $c$ in a given tile at
time $t+1$ is assumed to depend on the number of cells of other colors in
contiguous neighboring tiles at time $t$.} 
\label{fig:raw}
\end{figure}

Although tracking individual cell trajectories is difficult due to cell migration,
overlapping cells, changes in cell morphology, image artifacts, cell death and division, obtaining cell counts by cell type (represented by a certain color) is
straightforward and can be easily automated. To describe the spatial
distribution for different cell types, we propose to divide an image into a
number of contiguous regions (tiles) to form a regular lattice structure as
shown in Figure \ref{fig:raw} (a). We then record the frequency of
cells of different colors in each tile at subsequent time points,
 and based on which we model the spatial and temporal dependencies of the cell growth.

To model spatio-temporal data, one could choose to approximate the spatio-temporal process by a spatial process of time series, that is, to view the process as a multivariate spatial process where the multivariate dependencies are inherited from temporal dependencies. In other words, it can be seen as a temporal extension of spatial processes. 

The most popular way of developing a spatial process is through the conditionally auto-regressive (CAR) model proposed by \cite{besag1974spatial}. \cite{waller1997hierarchical} extend the CAR model into a spatio-temporal setting by allowing spatial effects to vary across time. However, the model lacks a specification of temporal dependency, as also noted by \cite{knorr1999bayesian}. More recently, \cite{quick2017multivariate} proposed a multivariate space-time CAR (MSTCAR) model, which is essentially a multivariate CAR model, where both temporal and between group dependencies are modelled as multivariate dependencies. Other works related to spatial process of time series include \cite{sans2008bayesian} and \cite{quick2016hierarchical}. 

Alternatively, one also think of the process as a time series of spatial process, or a spatial extension of time series. This is the approach we take in our spatio-temporal modelling. The underlying notion is that ``the temporal dependence is more natural to model than the spatial dependence" \citep{cressie2015statistics}. 

Following \cite{cox1981statistical}, it is useful to distinguish 
two modelling approaches for the analysis of time series data commonly seen in spatial-temporal modelling literature: the parameter-driven and observation-driven model.
In a parameter-driven model, the dependence between subsequent observations is modelled by a latent stochastic process, which evolves independently of the past history of the observation process. In contrast, in an observation-driven model, time dependence arises because the conditional expectation of the outcome given the past depends explicitly on the past values. 

For multivariate count data, the advantage of parameter-driven models is that one can easily assume that the conditional expectation of the observed process (on log-scale), as a latent process, is (multivariate) normal.  
There are extensive works related to latent spatio-temporal models under the Bayesian framework, including models with Gaussian data modelled by (multivariate) Gaussian process with an additive error (\cite{wikle1998hierarchical},  \cite{shaddick2002modelling}, \cite{bradley2015multivariate} and \cite{bradley2016multivariate}), Poisson data with conditional expectation modelled by  Gaussian latent process (\cite{mugglin2002hierarchical}, \cite{holan2015hierarchical} and Chapter 7 of \cite{cressie2015statistics}) and Poisson data with multivariate log-gamma latent process \citep{bradley2017computationally}. 
However, estimation of parameters in parameter-driven models requires considerable computational effort, as does prediction of the latent process. 

On the other hand, in observation-driven models, inference is possible in a (penalized) maximum likelihood framework and therefore can be easily fitted even for quite complex regression models \citep{davis2003observation}. 
\cite{schrodle2012assessing} proposed a parameter-driven spatio-temporal model and compared it with a similar observation-driven model proposed by \cite{paul2008multivariate}. They conclude that the parameter-driven models perform slightly better in terms of prediction in some cases, however, while the computation time for the observation-driven model is mostly less than a second, fitting a parameter-driven model takes several hours if it ever converges, because of the complexity with the latent autoregressive process. Besides, their model contains only five parameters, while in our application, the number of parameters of interest grows quadratically with the number of cell populations, which will make the parameter-driven models intractable even with a moderate number of cell populations. 

Therefore, we choose to work with a spatial extension of observation-driven time series. \cite{zeger1988markov} review various observation-driven time series models with a quasi-likelihood estimation.  \cite{fokianos2011log} develop and study 
the probabilistic properties of a log-linear autoregressive time series model for Poisson data, as an extension of the model considered by \cite{fokianos2009poisson}. See \cite{dunsmuir2015glarma} and \cite{kedem2005regression} for a  complete review. 

Literature about observation-driven spatio-temporal models, however, is relatively sparse. 
\cite{held2005statistical} propose a multivariate time series model where parameters are allowed to vary across space. \cite{paul2008multivariate} extended the model such that spatial dependences are captured by additional parameters that quantify the ``directed influence" of neighbouring areas at previous time points on the observation of interest.  \cite{paul2011predictive} further extend the model by introducing random effects. Note that these approaches model directly the conditional expectation of the count data, meaning they are using an identity link function, instead of the canonical log-link. Thus, it is required that the parameters are positive to ensure that the resulting conditional expectation is positive. \cite{knorr2003hierarchical} propose a space-time model for surveillance data, apart from separate seasonal and spatial components, they include an autoregressive term with a latent indicator.

In this paper, we develop a conditional spatial-temporal model for multivariate count data on tiled images, and provide its application on tiled images in the context of longitudinal cancer cell monitoring experiments. Our model enables us to measure the effect on the growth rate of each cell population and changes due to local cross-population interactions.
 Specifically, we consider a multivariate Poisson model with intensity modeled as a log-linear form similar to those in \cite{knorr2003hierarchical} and  \cite{fokianos2011log}, and we quantify spatio-temporal impacts of different cell populations in neighboring tiles through model parameters, as illustrated in Figure \ref{fig:raw} (b). 
Impacts are allowed to be positive or negative, and unlike those models that describe between group dependence through a covariance matrix, influences do not have to be symmetrical in our model. 
Another main advantage of the proposed framework is that it enables one to accommodate
spatio-temporal cell interactions for heterogeneous cell populations within a relatively parsimonious statistical model. 

Since the model complexity can be potentially very large in the presence of many cell types, it is also important to address the question of how to select an appropriate model by retaining only the meaningful spatio-temporal interactions between cell populations 
We cary out a model selection using the common model selection criteria for parametric models, the Akaike and the Bayesian information criteria (AIC and BIC).

The remainder of the paper is organized as follows. In Section
\ref{sec:methods}, we introduce the conditional spatio-temporal lattice model for
multivariate count data and develop  maximum likelihood inference tools. In
the same section, we discuss the asymptotic properties of our estimator and standard errors. In
Section \ref{sec:montecarlo}, we study the performance  of the new estimator
using simulated data. In Section \ref{sec:realdata}, we apply our method to analyze datasets from two in-vitro experiments: One where cancer cells are co-cultured with fibroblasts, and one where individually recognisable cloned cancer cell populations are cultured together in different combinations. In Section \ref{sec:conclusion}, we conclude and give final
remarks.

\section{Methods}\label{sec:methods}

\subsection{Multicolor spatial autoregressive model on the lattice} \label{sec:model}
 Let $\Lcal \in \mathbb{N}^2$ be a discrete lattice.
In the context of our application, the lattice is obtained by tiling a
microscope image into $n_\Lcal$ tiles, denoted by $\Lcal_n (\subset \Lcal)$.
The total number of tiles $n_\Lcal$ is a monotonically increasing function of $n.$
One can choose various forms of lattice, for example, the regular or hexagonal lattices. 
 For simplicity, we tile the image into $n \times n$ regular rectangular tiles, which makes $n_{\Lcal} =n^2.$
  An example of a tiled image with $n=10$ is shown in Figure
\ref{fig:raw} (a). 
Denote a pair of neighbouring tiles $\{i, j\}$ with $i \sim j$, if tiles $i$
 and $j$ share the same border or coincide ($i=j$).
Each tile may contain cells of different 
colors; thus, we let $\Ccal=\{ 1, \dots, \ncol\}$ be a finite set of colors and
denote by $\ncol$ the total number of 
colors. Let ${\pmb{Y}} = \{ {\pmb Y}_{t}, t  = 1, \dots,T\}$ be the sample of
observations where ${\pmb{Y}}_t = \{ {\pmb{Y}}_t^{(c)}, c \in \Ccal \}$ is the collection of observations at time point $t$, and 
${\pmb{Y}}^{(c)}_{t}=(Y_{1,t}^{(c)}, \dots, Y_{n_\Lcal,t}^{(c)})^\top$ is the vector of observed 
frequencies for color $c$ on the lattice $\Lcal_n$ at time $t$. The 
joint distribution for the spatio-temporal process on the  lattice is difficult to
specify, due to local spatial 
interactions for neighboring tiles and global interactions occurring at the level of the entire image. An additional issue is that cells tend to be clustered together due to the cell division process and other biological
mechanisms; thus it is not uncommon to observe low counts in a considerable portion of tiles. 
In typical longitudinal experiments, the number of time points seldom go beyond $50$ due to experimental, storage and processing cost, while $n_\Lcal$ can be relatively large. So we work under the framework where $T$ is assumed to be finite, while $n_\Lcal$ is allowed to grow to infinity. 

We suppose 
that the count for the $i$th tile
$Y^{(c)}_{i,t}$
follows a marginal Poisson distribution $Y^{(c)}_{i,t}|{\pmb{Y}}_{t-1}  \sim
\text{Pois}(\lambda^{(c)}_{i,t})$, with  intensity  modeled by the canonical
log-link $v^{(c)}_{i,t} =\log
\lambda_{i,t}^{(c)} 
$, where $v^{(c)}_{i,t}$ takes the following spatial autoregressive form:
\begin{align}
v^{(c)}_{i,t}  & =   \alpha^{(c)} + \sum_{c' \in \Ccal}  \beta^{(c|c')} 
S_{i,t-1}^{(c')} , \label{intensity1} \\
S_{i,t-1}^{(c')}  &  = \dfrac{1}{n_i} \sum_{\substack{i\sim j: \
j\in \Lcal_n}} \log\left( 1+  Y_{j,t-1}^{(c')} \right), \label{intensity2}
\end{align}
for all $c \in \Ccal,  t = 1, \dots, T$, with $n_i = \{ \# j: i \sim j, j\in \Lcal_n\}$ being the number of tiles in a
neighbourhood of tile $i$. 
Although we are adopting the regular grids for simplicity, the model is readily applicable to other tiling strategies. Changing the tiling strategy would only change the realisations of $S_{i,t-1}^{(c')}$ in (2).

Here, we assume that the conditional count for different tiles at time $t$ is independent conditioning on information from $t-1$, i.e. 
$$
P\big(Y^{(c)}_{i,t} Y^{(c')}_{j,t}| {\pmb{Y}}_{t-1} \big) = P\big(Y^{(c)}_{i,t}| {\pmb{Y}}_{t-1} \big) P\big(Y^{(c')}_{j,t}| {\pmb{Y}}_{t-1} \big), 
$$ 
for all $c,c' \in \Ccal,  t = 1, \dots, T,$ and $ i,j \in \Lcal_n, i \neq j.$
This does not suggest that they ($Y^{(c)}_{i,t} $ and $Y^{(c')}_{j,t}$) are independent, but rather that their spatio-temporal dependence is due to the structure of intensity $\lambda_{i,t}^{(c)}$ in (\ref{intensity1}). Conditional independence is a commonly used assumption for spatio-temporal models in a non-gaussian setting \cite{waller1997hierarchical,wikle2003climatological}, since it's exceedingly difficult to work with multivariate non-Gaussian distribution \cite{cressie2015statistics}.

 The elements of the parameter vector ${\pmb{\alpha}} = (
\alpha^{(1)},\dots, \alpha^{(\ncol)})^\top$ are 
main effects corresponding to a baseline average count for cells of different
colors.  The spatio-temporal interactions are measured by the statistic
$S_{i,t-1}^{(c')}$ in (\ref{intensity2}), which essentially counts the number of 
cells of color 
$c'$ in the neighborhood of tile $i$ at time $t-1$. Hence, the
autoregressive parameter $\beta^{(c|c')}$ is interpreted as positive or
negative 
change in the average number of cells with color $c$, due to interactions with
cells of color $c'$ in neighbouring tiles. A positive (or a negative) sign of 
$\beta^{(c|c')}$ means that the presence of cells of color $c'$ in neighboring
tiles promotes (or inhibits) the growth of cells of color $c$. The 
spatio-temporal effects $\beta^{c|c'}, c,c' \in \Ccal ,$ are collected in the $\ncol \times \ncol$ weighted
incidence matrix ${\pmb{\mathcal{B}}}$. This may be used
to generate weighted directed
graphs, as shown in the example of Figure \ref{f:realdata2}, where the nodes of the directed graph correspond to
cell  types, and the directed edges are negative or positive spatio-temporal
interactions between cell types.

 Equation (\ref{intensity1}) could be extended to some more specific form, for example,  $v^{(c)}_{i,t}   =   \alpha^{(c)} + \sum_{c' \in \Ccal}  \beta_1^{(c|c')}  S_{i,t-1}^{(c')} + \beta_0^{(c|c')} \log\left( 1+  Y_{i,t-1}^{(c')} \right)$, where $ \beta_1^{(c|c')}$ are interpreted as the effect of cells of color $c'$ from neighbouring (but not the same) tiles have on the growth of cells with color $c$, while $ \beta_0^{(c|c')}$ as the effect of cells of color $c'$ from the same tile. However, we stick to the model in (\ref{intensity1}) because we have no evidence showing that the more complex model is advantageous from model selection view point. 

 We choose to work with a log-linear form for the autoregressive equation of $v^{(c)}_{i,t}$ in Equation (\ref{intensity1}), where we apply a logarithmic transform and add $1$ to the counts at time $t-1$, $Y^{(c)}_{i,t-1}$.
 It offers several advantages compared to the more commonly used linear form. 
First, $\lambda^{(c)}_{i,t}$ and $Y^{(c)}_{i,t-1}$ are transformed on the same scale.
Moreover, this model can accommodate both positive and negative correlations, while it is not possible to account for positive association in a stationary model if past counts are directly included as explanatory variables. 
For example, with the model $v_{i,t} = \alpha + \beta Y_{i,t-1}$ for a single color, 
the intensity would be $\lambda_{i,t} = \exp\left(\alpha \right) \exp\left(\beta  Y_{i,t-1} \right),$
which may lead to instability of the Poisson means if $\beta>0$ since $\lambda_{i,t}$ is allowed to increase exponentially fast. 
Finally, adding $1$ to $Y^{(c)}_{i,t-1}$ is for coping with zero data values, since $\log(Y^{(c)}_{i,t-1})$ is not defined when $Y^{(c)}_{i,t-1} = 0$, which arises often, and it maps zeros of $Y^{(c)}_{i,t-1}$ into zeros of $\log(1+ Y^{(c)}_{i,t-1})$.

\subsection{Likelihood inference} \label{sec:clinference}

Let ${\pmb{\theta}} $
be the overall parameter vector 
${\pmb{\theta}} = ({\pmb{\alpha}} ^\top, \text{vec}({\pmb{ \mathcal{B}}})^\top )^\top \in \mathbb{R}^{p}$,  
where ${\pmb{\alpha}} $ is a $n_\Ccal$-dimensional vector defined in Section \ref{sec:model} and 
${\pmb{\mathcal{B}}}$ is a $n_\Ccal \times n_\Ccal$ matrix of colour interaction effects,  
 $p = \ncol(1+\ncol)$ is the total number of parameters. In this section, 
we develop a weighted maximum likelihood estimator for our model,
\begin{equation}\label{CL}
L_n({\pmb{\theta}} ) =   \prod_{t=1}^T  \prod_{c \in\Ccal} \prod_{i \in \Lcal_n}  P(Y_{i,t}^{(c)} |{\pmb{Y}} _{t-1};  {\pmb{\theta}}  )   =   
  \prod_{t=1}^T  \prod_{c \in \Ccal} \prod_{i \in \Lcal_n}  \Bigg( e^{-\lambda_{i,t}^{(c)}({\pmb{\theta}} )} 
  \dfrac{{\lambda_{i,t}^{(c)}({\pmb{\theta}} )}^{y_{i,t}^{(c)}}}{y_{i,t}^{(c)}!}  \Bigg),
 \end{equation}
 where $\lambda_{i,t}^{(c)}({\pmb{\theta}} )$ is the expected number of cells with color 
 $c$ in tile $i$ at time $t$, defined in (\ref{intensity1}). 
The maximum likelihood estimator (MLE), $\hat{{\pmb{\theta}} }$, is
obtained by maximizing the weighted log-likelihood function
 \begin{equation}\label{cl}
\ell_n({\pmb{\theta}} ) =  \sum_{i \in \Lcal_n} \sum_{t = 1}^T  \sum_{c \in
\Ccal} \left[ Y_{i,t}^{(c)} v^{(c)}_{i,t}({\pmb{\theta}} ) - 
\exp \left\{ v^{(c)}_{i,t}({\pmb{\theta}} ) \right\} \right], 
 \end{equation}
where $v_{i,t}^{(c)}({\pmb{\theta}} ) \equiv \log \lambda_{i,t}^{(c)}({\pmb{\theta}} )$. 
Equivalently, $\hat{{\pmb{\theta}} }$  is formed by solving the weighted estimating equations
\begin{equation}\label{score}
0 = {\pmb{u}}_n({\pmb{\theta}} ) \equiv \dfrac{1}{n_\Lcal}\nabla \ell_n({\pmb{\theta}} ) = \dfrac{1}{n_\Lcal}   \sum_{i \in \Lcal_n} \sum_{t=1}^T   {\pmb{\gamma}}_{i,t}({\pmb{\theta}} ) \otimes \nabla {\pmb{v}}_{i,t},
\end{equation}
where $ {\pmb{\gamma}}_{i,t}({\pmb{\theta}} ) = \left( y^{(1)}_{i,t} - \exp \left\{ v^{(1)}_{i,t}({\pmb{\theta}} ) \right\}, \dots,y^{(n_\Ccal)}_{i,t} - \exp \left\{ v^{(n_\Ccal)}_{i,t}({\pmb{\theta}} ) \right\} \right)$, $\otimes$ denotes the Kronecker product, 
 $\nabla$ is the gradient operator with respect to ${\pmb{\theta}} $ 
and $\nabla {\pmb{v}}_{i,t} \equiv \nabla {\pmb{v}} ^{(c)}_{i,t}({\pmb{\theta}} ) = (  1, S_{i,t-1}^{(1)}, \dots,
S_{i,t-1}^{(\ncol)})^\top$.  

Our empirical results show that this choice performs reasonably well in terms of estimation accuracy in all our numerical examples
 and guarantees optimal variance for the estimator $\hat{{\pmb{\theta}} }$ under correct model specification.
The solution to Equation (\ref{score}) is obtained by a standard Fisher scoring
algorithm, which is found to be stable and converges fast in all our numerical
examples.

Finally, in practical applications it is also important to address the question of how to select an appropriate model by retaining only the meaningful spatio-temporal interactions between cell populations, and avoid over-parametrized models.  Model selection plays an important role by  balancing goodness-of-fit and model complexity.  Here, we select non-zero model parameters based traditional model selection approaches: the 
Akaike Information criterion, $AIC = -2\ell(\hat{{\pmb{\theta}} }) + 2p$, and the Bayesian information criterion, $BIC = -2\ell(\hat{{\pmb{\theta}} }) + p\log(|n_\Lcal T|)$.

 \subsection{Asymptotic properties and standard errors} \label{sec:se}
 
In this section, we overview the asymptotic behavior of the estimator introduced in Section \ref{sec:clinference}. In our setting we consider a fixed number of time points, $T$, whilst the lattice $\Lcal_n$ is allowed to increase. This reflects the notion that the statistician is allowed to  choose an increasingly fine tiling grid as the number of cells increases.  If the regularity conditions stated in the Appendix hold, then  $\sqrt{n_\Lcal}{\pmb{ H}}_n({\pmb{\theta}} _0)^{1/2}(\hat{{\pmb{\theta}} }_n - {\pmb{\theta}} _0)$ converges in distribution to a $p$-variate normal distribution with zero mean vector and identity variance, as $n_\Lcal \rightarrow \infty$, with  ${\pmb{H}} _n({\pmb{\theta}} )$ given in (\ref{eq:H}).  Asymptotic normality of  $\hat{{\pmb{\theta}} }_n$ follows by applying the limit theorems for M-estimators for nonlinear spatial models developed by \cite{jenish2009central}. One condition required to ensure this behavior is that ${\pmb{Y}} _t$ has constant entries at the initial time point $t=0$, which is quite realistic since typically cells are seeded randomly at the beginning of the experiment. Our proofs mostly check  $\alpha$-mixing conditions and $\Lcal_2$-Uniform Integrability of the score functions ${\pmb{u}} _{i,t}({\pmb{\theta}} )$ ensures a pointwise law of large numbers, with additional stochastic equicontinuity, a uniform version of the law of large numbers required by \cite{jenish2009central}.

The asymptotic asymptotic variance of $\hat {\pmb{\theta}} $ is ${\pmb{V}} _n(\hat{{\pmb{\theta}} })= {\pmb{H}} _n^{-1}({\pmb{\theta}} _0)$, where 
${\pmb{H}} _n({\pmb{\theta}} )$ is the $p \times p$  Hessian matrix
\begin{align} \label{eq:H}
{\pmb{H}} _n({\pmb{\theta}} ) = - E \left[ \nabla^2  \ell({\pmb{\theta}} )\right] = - E \left(
 \sum_{i \in \Lcal_n}
\nabla {\pmb{u}} _i({\pmb{\theta}} ) \right),
\end{align}
with 
$ {\pmb{u}} _i({\pmb{\theta}} ) = {\pmb{u}}_{i,1}({\pmb{\theta}} ) + \cdots + {\pmb{u}}_{i,T}({\pmb{\theta}} )$ being the partial
score function for the $i$th tile. Direct evaluation of ${\pmb{H}}({\pmb{\theta}})$ may be challenging since the expectations in (\ref{eq:H}) is intractable. Thus, we estimate ${\pmb{H}}_n({\pmb{\theta}})$ by the empirical counterpart 
$$
\hat{{\pmb{H}}}_n({\pmb{\theta}}) = 
 \begin{pmatrix}
 \hat{{\pmb{H}}}^{(1)}({\pmb{\theta}}) & {\pmb{0}} & \cdots & {\pmb{0}} \\
 {\pmb{0}}  &  \hat{{\pmb{H}}}^{(2)}({\pmb{\theta}}) & \cdots & {\pmb{0}}  \\
  \vdots  & \vdots  & \ddots & \vdots  \\
  {\pmb{0}}  & {\pmb{0}}  & \cdots &  \hat{{\pmb{H}}}^{(n_\Ccal)}({\pmb{\theta}}) 
 \end{pmatrix},
$$
where 
\begin{align}\label{hatH}
\hat{{\pmb{H}}}^{(c)}({\pmb{\theta}}) & = \sum_{i \in
\Lcal_n} \sum_{t=1}^T 
\exp\left[v_{i,t}^{(c)}({\pmb{\theta}})\right] \left[ \nabla
{\pmb{v}}_{i,t}\right]\left[
\nabla {\pmb{v}}_{i,t}\right]^\top. 
\end{align}
Note that the above estimators approximate the quantities in
formula (\ref{eq:H}) by conditional expectations. 
Our numerical results suggest that the above variance approximation yields 
confidence intervals 
with  coverage  very close to the nominal level $(1-\alpha)$.  Besides the above formulas, we also consider confidence intervals obtained by a parametric 
bootstrap approach. Specifically, we generate $B$ bootstrap samples ${\pmb{Y}}^*_{(1)}, \dots, {\pmb{Y}}^*_{(B)} $ by sampling at subsequent times from the 
conditional model specified in Equations (\ref{intensity1}) and (\ref{intensity2}) with ${\pmb{\theta}} = \hat{{\pmb{\theta}}}$. From such bootstrap samples, we obtain bootstrapped estimators, 
$\hat{{\pmb{\theta}}}^*_{(1)}, \dots, \hat{{\pmb{\theta}}}^*_{(B)}$, which are used to estimate $var(\hat {\pmb{\theta}}_0)$ by the usual covariance estimator $\hat{{\pmb{V}}}_{boot} (\hat{{\pmb{\theta}}}) = \sum_{b=1}^B (\hat{{\pmb{\theta}}}^*_{(b)} - {\overline{{\pmb{\theta}}} }^* )^2/ (B-1)$, where ${\overline{{\pmb{\theta}}} }^* =\sum_{b=1}^B\hat{{\pmb{\theta}}}^*_{(b)}/B $. Finally, a $(1-\alpha) 100 \%$ confidence interval for ${\pmb{\theta}}_j$ is obtained as 
$\hat{{\pmb{\theta}}}_j \pm z_{1-\alpha/2}  \{\hat{{\pmb{V}}}\}_{jj}^{1/2}$, where 
$z_q$ is the $q$-quantile of a standard normal distribution, and $\hat{{\pmb{V}}}$ is an estimate of $var(\hat {\pmb{\theta}})$ obtained by either Equation (\ref{hatH}) or   bootstrap resampling.

\section{Monte Carlo simulations} \label{sec:montecarlo}

In our Monte Carlo experiments, 
we generate data from a Poisson model as follows. 
At time $t=0$, we populate $n_{\Lcal}$ tiles using equal counts for cells of different colors. 
For $t = 1, \dots, T$, observations are drawn from the multivariate Poisson model
$
Y^{(c)}_{i,t}|{\pmb{Y}}_{t-1} \sim \text{Poisson}(\lambda^{(c)}_{i,t}),  c \in \Ccal.
$
Recall that the rate $\lambda^{(c)}_{i,t}$ defined in Section \ref{sec:model} contains autoregressive coefficients $\beta^{(c|c')}$, which are collected in the $n_\Ccal \times n_\Ccal $ matrix $\mathcal{B}$.

We assess the performance of MLE under different settings concerning the size and sparsity of  $\mathcal{B}$. Consider the three models with the following choices of $\mathcal{B}$:  

\[ {\pmb{\mathcal{B}_1}} = \left( \begin{array}{ccc}
0.7 & -0.7 & 0.7 \\
0.7 & 0.7 & -0.7 \\
-0.7 & 0.7 & 0.7\end{array} \right), 
  {\pmb{\mathcal{B}_2}} = \left( \begin{array}{ccc}
0.05 & -0.15 & 0.25 \\
0.35 & 0.45 & -0.55 \\
-0.65 & 0.75 & 0.85 \end{array} \right), 
  {\pmb{\mathcal{B}_3}} = \left( \begin{array}{ccc}
0.7 & -0.7 & 0.7 \\
0 & 0.7 & 0 \\
0 & 0 & 0.7\end{array} \right).   \]

Denote Model $i$ as the model corresponding to $\mathcal{B}_i, i = 1,2,3$. In Model 1, all the effects in ${\pmb{\mathcal{B}}}$ have the same size; in Model 2, the effects have decreasing sizes; Model 3 is the same as Model 1, but with some interactions exactly equal to zero. 

We set $\alpha^{(1)}=  \cdots = \alpha^{(n_{\Ccal})} = -0.1$ for all three models, which 
The above parameter choices reflect the situation where the generated process ${\pmb{Y}}$ has a moderate growth. 

In Tables \ref{t:Sim1} and \ref{t:Sim2}, we show results based on 1000 Monte Carlo runs generated from Models 1-3, for $n=25, n_\Ccal = 3$ and $T=10$ and $25$. 
In Table \ref{t:Sim1}, we show Monte Carlo estimates of squared bias and variance of $\hat{{\pmb{\theta}}}$. 
Both squared bias and variance of our estimator are quite small in all three models, and decrease as $T$ gets larger. 
The variances of Model 2 are slightly larger than those in the other two models due to the increasing difficulty in estimating parameters close to zero.  

\begin{table}[h] 
\begin{center}
 \begin{tabular}{ccccccccccc }\toprule
& \multicolumn{2}{c}{$T = 10$} & \phantom{abc}& \multicolumn{2}{c}{$T = 25$} &
\phantom{abc}& 
\\ \cmidrule{2-3} \cmidrule{5-6} 
 & $\widehat{\text{Bias}}^2$ & $\widehat{\text{Var}}$ && $\widehat{\text{Bias}}^2$ & $\widehat{\text{Var}}$ \\ \midrule
Model 1 & 0.45(0.57) & 5.75(0.26)  & &0.29(0.32) & 2.36(0.11)\\
Model 2 & 0.64(0.91) & 9.66(0.42) & &0.67(0.71) & 4.45(0.20)\\ 
Model 3 & 0.77(0.97) & 8.09(0.36) && 0.52(0.51) & 3.47(0.16)\\  \bottomrule
\end{tabular} 
\end{center} 
\caption{Monte Carlo estimates for squared bias $(\times 10^{-6})$ and variance $(\times 10^{-4})$ of the MCLE for three models with time points $T= 10, 25.$ Simulation standard errors are shown in parenthesis. The three models differ in terms of the coefficients $\beta^{(c|c')}, c,c' \in \Ccal$, as described in Section \ref{sec:montecarlo}: Non-zero equal effects (Model 1), non-zero decreasing interactions (Model 2), and sparse effects (Model 3). For all models, $ \alpha^{(c)} = -0.1, c = 1, 2, 3$. Estimates are based on 1000 Monte Carlo runs.}\label{t:Sim1}
\end{table}

 In Table \ref{t:Sim2}, we report the coverage probability for symmetric confidence intervals of the form $\hat 
{{\pmb{\theta}}} \pm z_{1-\alpha/2} {\widehat{sd}(\hat{{\pmb{\theta}}}) }$, where $z_q$ is the $q-$quantile for a standard normal distribution, with $\alpha = 0.01, 0.05, 0.10.$
The standard error, $\widehat{sd}(\hat{{\pmb{\theta}}})$, is obtained by the sandwich and the parametric bootstrap estimate, $\hat{{\pmb{V}}}_{est}$ and $\hat{{\pmb{V}}}_{boot}$, described in Section \ref{sec:se}.
The coverage probability of the confidence intervals are very close to the nominal level for both methods. 
\begin{table}[h] 
\begin{center}
 \begin{tabular}{ccccccccccccccccccc }\toprule
& \phantom{abc}& \multicolumn{2}{c}{$T = 10$} & \phantom{abc}& \multicolumn{2}{c}{$T = 25$} &
\phantom{abc}
\phantom{abc} \\ \cmidrule{3-4} \cmidrule{6-7} 
 && $\hat{{\pmb{V}}}_{boot}$ & $\hat{{\pmb{V}}}_{est}$ && $\hat{{\pmb{V}}}_{boot}$ & $\hat{{\pmb{V}}}_{est}$    \\ \midrule
&Model 1 &98.6  &99.0 & &98.9 & 99.0 \\
$\alpha = 0.01$ &Model 2 & 99.0 & 99.0  &&98.8 &98.9 \\ 
&Model 3 & 98.9 &99.0 && 98.9 & 98.9\\  \midrule
&Model 1 & 94.2  & 95.2  && 94.9 & 95.0 \\
$\alpha = 0.05$ &Model 2 & 95.2 &95.1  & &95.0 & 95.3 \\ 
&Model 3 & 95.4 &95.5 && 94.9 & 95.1\\  \midrule
&Model 1 & 89.2  & 90.3 && 90.1 &90.3 \\
 $\alpha = 0.10$&Model 2 & 90.6 &90.0  &&89.7 &90.0\\ 
&Model 3 & 90.6 & 90.6 && 90.2 &90.2 \\ \bottomrule
\end{tabular} 
\end{center} 
\caption{ Monte Carlo estimates for the coverage probability of $(1-\alpha)\%$ confidence intervals
 $\hat {{\pmb{\theta}}} \pm z_{1-\alpha/2} {\widehat{sd} (\hat{{\pmb{\theta}}})}$, with ${\widehat{sd}(\hat{{\pmb{\theta}}}) }$ obtained using bootstrap ($\hat{{\pmb{V}}}_{boot}$) and sandwich ($\hat{{\pmb{V}}}_{est}$) estimators in Section \ref{sec:methods} and \ref{sec:montecarlo}. The three models differ in terms of the coefficients $\beta^{(c|c')}, c,c' \in \Ccal$ as described in Section \ref{sec:montecarlo}: Non-zero equal effects (Model 1), non-zero decreasing interactions (Model 2), and sparse effects (Model 3). For all models, $ \alpha^{(c)} = -0.1, c = 1, 2, 3$, estimates are based on 1000 Monte Carlo runs.}\label{t:Sim2}
\end{table}

In Table \ref{t:Sim3}, we show results for the model selection based on 1000 Monte Carlo samples from Model 3 using the AIC  and the BIC given in Section 2 for $n=25$ and $T=10, 25$. 
We report Type A error  (a term is not selected when it actually belongs to the true model ) and Type B error (a term is selected when it is not in the true model ).
 For both AIC and BIC model selection is more accurate for large $T$.
As expected AIC tends to over select, and BIC outperforms AIC, 
 with zero Type A error,  and very low Type B error.

 \begin{table}[h] 
\begin{center}
 \begin{tabular}{cccccccccccccccccccccccccccc }\toprule
& \multicolumn{3}{c}{$T = 10$} &\multicolumn{3}{c}{$T = 25$}  \\ 
\cmidrule{2-3} \cmidrule{5-6}  
& Type A  & Type B  &  &  Type A & Type B & \\ \midrule
AIC  & 0.00  & 10.00  & &  0.00  & 10.38 &&\\ \midrule
BIC  & 0.00 &  0.22   &  &0.00 & 0.20  &\\ \bottomrule
\end{tabular} 
\end{center} 
	\caption{ Monte Carlo estimates for $\%$ Type A error (a term is not selected when it actually belongs to the true model) and $\%$ Type B error (a term is selected when it is not in the true model) 
using AIC and BIC criteria. 
Results are based on 1000 Monte Carlo samples generated from Model 3 with $n=25$ and $T=10, 25$. }\label{t:Sim3}
\end{table}

\section{Analysis of the cancer cell growth data} \label{sec:realdata}

Cancer cell behavior is believed to be determined by several factors including genetic profile and differentiation state. However, the presence of other cancer cells and non-cancer cells has also been shown to have a great impact on overall tumor behavior \cite{tabassum2015tumorigenesis,kalluri2006fibroblasts}. It is therefore important to be able to dissect and quantify these interactions in complex culture systems. The data sets in this section represent two scenarios: 
 cancer cell-fibroblast co-culture and cloned cancer cell co-culture experiments. 
The data sets analyzed consist of counts of cell types (different cancer cell populations expressing different fluorescent proteins, and non-fluorescent fibroblasts) from 9 subsequent images taken at an 8-hour frequency over a period of 3 days using the Operetta high-content imager (Perkin Elmer). Information regarding cell type (fluorescent profile) and spatial coordinates for each individual cell were extracted using the associated software (Harmony, Perkin Elmer). Each image was subsequently tiled using a $25\times 25$ regular grid.

\subsection{Cancer cell-fibroblast co-culture experiment}
In this experiment,  cancer cells are co-cultured with fibroblasts, a predominant cell type in the tumor microenvironment, believed to affect tumor progression, partly due to interactions with and activation by cancer cells \cite{kalluri2006fibroblasts}. 
In this experiment, fibroblasts (F) are non-fluorescent whereas cancer cells fluoresce either in the red (R) or green (G) channels due to the experimental expression of mCherry  or GFP proteins, respectively. Cells were initially seeded at a ratio of 1:1:2 (R:G:F).

\begin{figure}[ht] 
\centering
\includegraphics[scale = 0.9]{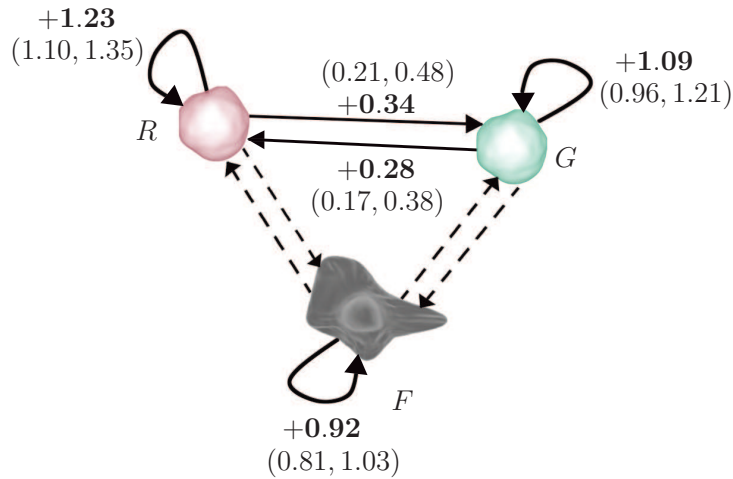}
\caption{Directed graph showing fitted spatio-temporal interactions between GFP cancer cells (G), mCherry cancer cells  (R) and fibroblasts (F).
The solid and dashed arrows represent respectively the significant and not significant interactions between cell types at the $95\%$ confidence level. 
 }
\label{f:realdata2}
\end{figure}

\paragraph{Model selection and inference.} We applied our methodology
to quantify the magnitude and direction  of the impacts have on growth for the considered cell types. 
To select the relevant terms in the intensity expression (\ref{intensity1}), we carry out model selection using the BIC model selection criterion.
In Table \ref{t:Real}, we show estimated parameters for the full and the BIC models, with bootstrap $95\%$ confidence intervals in parenthesis. 
Figure \ref{f:realdata2} illustrates estimated spatio-temporal impacts between cell types using a directed graph. The solid and dashed arrows represent respectively significant and not significant impacts between cell types at the $95\%$ confidence level. 
Significant impacts coincide with parameters selected by BIC. 

The interactions within each cell type ($\hat{\beta}^{(c|c)}, c= R, G, F$) are significant, which is consistent with healthy growing cells.
As anticipated, the effects $\hat{\beta}^{(c|c)}$ for the cancer cells are larger
than those for the slower growing fibroblasts. The validity of the estimated
parameters is also supported by the similar sizes of the parameters for the green and red cancer cells. 
This is expected, since the red and green cancer cells are biologically identical except for the
fluorescent protein they express.
Interestingly, the size of the estimated effects within both types of cancer cells ($\hat{\beta}^{(c|c)}, c= R, G$)
 are larger than  the impact they have on one another ($\hat{\beta}^{(G|R)}$ and $\hat{\beta}^{(R|G)}$). This is not surprising, since $\hat{\beta}^{(c|c)} (c= R, G)$
reflects not only  impacts between cells from the same cell population, but also cell proliferation. The fact that we are able to detect the impacts between the red and green cancer cells confirms that our methodology is sensitive enough to detect biologically relevant impacts even though no interactions were found between the cancer cells and the fibroblasts. This might be due to the fact that we used normal fibroblasts that had not previously been in contact with cancer cells and thus had not been activated to support tumor progression as is the case with cancer-activated fibroblasts.

 \begin{table}[h] 
\begin{center}
\scalebox{0.9}{
 \begin{tabular}{ccccccccc }\toprule\label{Real}
& \multicolumn{3}{c}{Full model} & \phantom{abc}
\phantom{abc} \\ \cmidrule{2-4} 
$c=$ & $G$ & $R$ & $F$ \\ \midrule
$\widehat{\alpha}^{(c)}$ &  \;-0.99 (-1.19, -0.79) & \;-0.50 (-0.70, -0.30) & \;-0.26 (-0.45, -0.06)   \\
$\widehat{\beta}^{(G|c)}$ & 1.23 (1.10, 1.35) & 0.34 (0.21, 0.48) & 0.12 (-0.03, 0.27) \\ 
$\widehat{\beta}^{(R|c)}$ & 0.28 (0.17, 0.38) & 1.09 (0.96, 1.21) & \;0.02 (-0.09, 0.13) \\  
$\widehat{\beta}^{(F|c)}$ & 0.10 (-0.01, 0.21) & \;0.02 (-0.07, 0.12) & 0.92 (0.81, 1.03)  \\  \midrule
& \multicolumn{3}{c}{BIC model} & \phantom{abc}
\phantom{abc} \\ \cmidrule{2-4} 
$c=$ & $G$ & $R$ & $F$\\ \midrule
$\widehat{\alpha}^{(c)}$ &\;-0.88 (-1.04, -0.72)& \;-0.49 (-0.66, -0.31)& \;-0.19 (-0.36, -0.02) \\
$\widehat{\beta}^{(G|c)}$ & 1.24 (1.11, 1.37)& 0.35 (0.21, 0.48)& /\\ 
$\widehat{\beta}^{(R|c)}$ & 0.28 (0.17, 0.39)& 1.09 (0.96, 1.21)& /\\  
$\widehat{\beta}^{(F|c)}$ & /& /& 0.93 (0.82, 1.04) \\ 
\bottomrule
\end{tabular} }
\end{center} 
\caption{Estimated parameters for the full and the BIC models based on the cancer cell growth data described in Section \ref{sec:realdata}. 
Bootstrap $95\%$ confidence intervals based on $50$ bootstrap samples are given in parenthesis. }
\label{t:Real}
\end{table}

\paragraph{Goodness-of-fit and one-step ahead prediction}

\begin{figure}[ht]
\centering
\includegraphics[scale = 0.9]{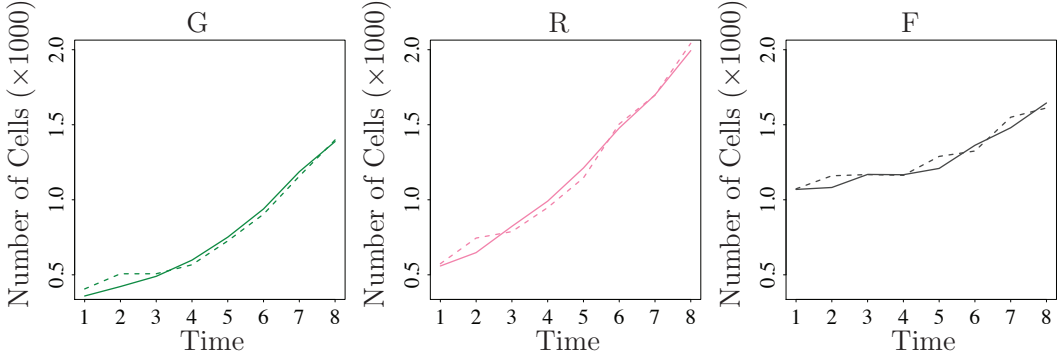}
\caption{ Goodness-of-fit of the estimated model. 
Observed (solid) and predicted (dashed) number of GFP cancer cells (G), mCherry cancer cells (R) cancer cells and fibroblasts (F) for the entire image. 
 Predicted cell counts for each cell type in each tile $\hat{y}_{i,t}^{{(c)}}$ is generated from the conditional Poisson model with intensity $\hat{\lambda}_{i,t}^{{(c)}}$
 defined in Equation (\ref{intensity1}) and (\ref{intensity2}), where the coefficients $\hat{\beta}^{(c|c')}$ are estimated from the entire dataset.} 
  \label{f:goodnesofffit}
\end{figure}

To illustrate the goodness-of-fit of the estimated model, we generate cell counts for each type in each tile, $\hat{y}_{i,t}^{{(c)}}$, from the Pois($\hat{\lambda}_{i,t}^{{(c)}}$) distribution for $t \leq 1$, where $\hat{\lambda}_{i,t}^{{(c)}}$ is computed using observations at time $t-1$, with parameters estimated from the entire dataset.   
In Figure \ref{f:goodnesofffit}, we compare the actually observed and generated cell counts for GFP cancer cells (G) and mCherry cancer cells (R) and fibroblasts (F) across the entire image. 
The solid and dashed curves for all cell types are close, suggesting that the model fits the data reasonably well. 
As anticipated, the overall growth rate for the red and green cancer cells are similar, and sensibly larger than the growth rate for fibroblasts. 

\begin{figure}[ht] 
\centering
\includegraphics[scale = 0.9]{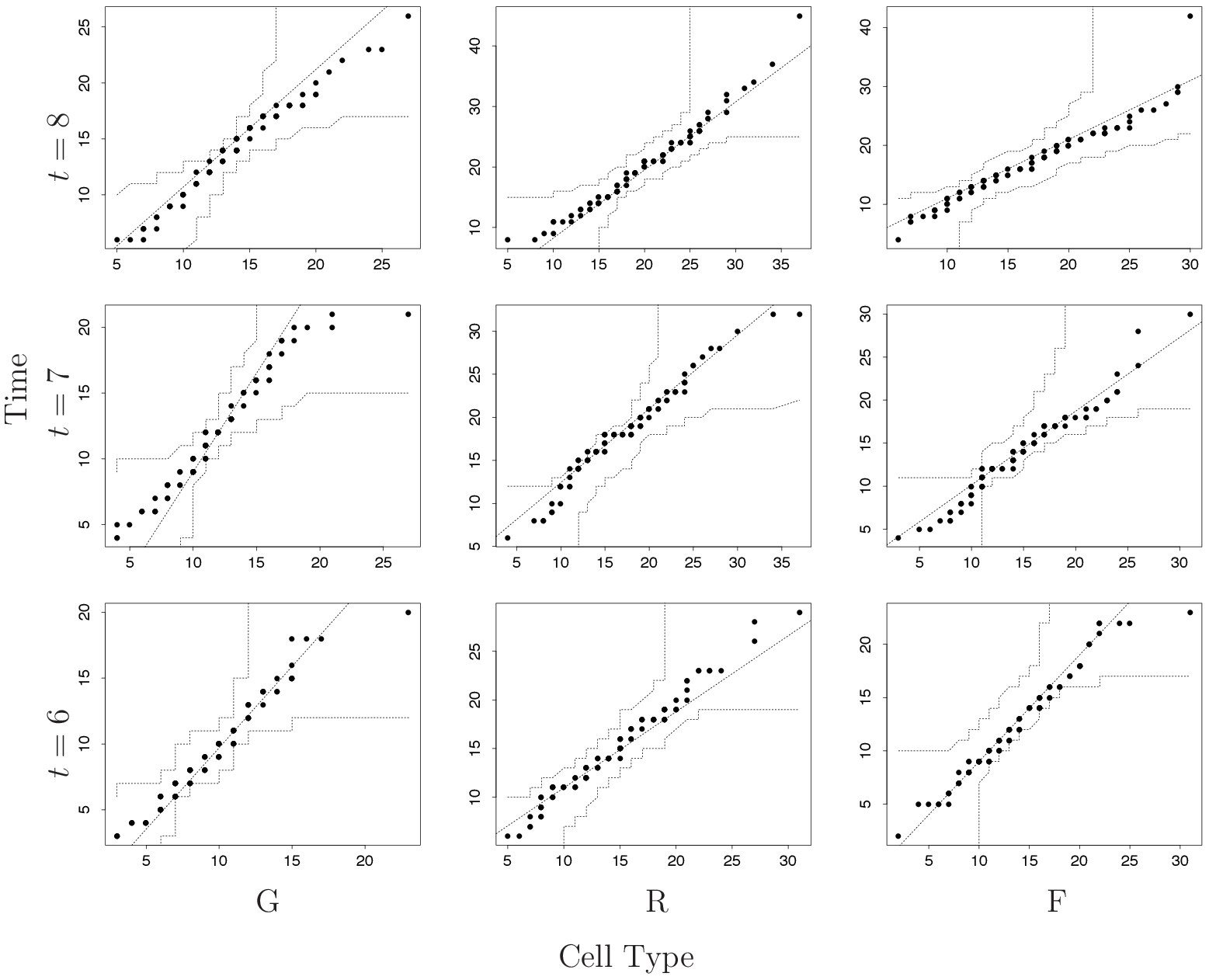}
\caption{ QQ-plots for cell growth, comparing observed (horizontal axis) and one-time ahead predicted (vertical axis) cell counts per tile on the entire image at times $t=6,7,8$ for GFP cancer cells (G), mCherry cancer cells (R) and fibroblasts (F). 
One-time ahead predictions are based on the model fitted using a moving window of five time points. 
}\label{qqplot2}
\end{figure}

To assess the prediction performance of our method, we consider one-step-ahead forecasting using parameters estimated from a moving window of five time points. 
In Figure \ref{qqplot2}, we show quantiles of observed cell counts against predicted counts for each tile. The upper and lower  $95\%$ confidence bounds are computed non-parametrically by taking 
$\hat{F}^{-1}_1  \big(\hat{F}_0 (y_{t}^{(c)})  - 0.95 \big)$ and $ \hat{F}^{-1}_1  \big(\hat{F}_0 (y_{t}^{(c)})  + 0.95 \big)$, where $\hat{F}_0$ and $\hat{F}_1$ are the empirical distributions of the observations and predictions at time $t$ respectively \citep{koenker2005quantile}.
The identity line falls within the confidence bands in each plot, indicating a satisfactory prediction performance. 

\section{Conclusion and final remarks}\label{sec:conclusion}

In this paper, we introduced a conditional spatial autoregressive model and accompanying inference tools 
for multivariate spatio-temporal cell count data. The new methodology enables one to measure the overall cell 
growth rate in longitudinal experiments and spatio-temporal interactions with either homogeneous or heterogeneous cell populations. 
The proposed inference approach is computationally
tractable and strikes a good balance between computational feasibility and statistical accuracy. Numerical findings from simulated and real data 
in Sections 3 and 4 confirm the validity of the proposed  approach in terms of prediction, goodness-of-fit and estimation accuracy.

The data sets described in this paper serve as a proof-of-concept that the proposed methodology works. However, the potential applications and the relevant questions that the methodology can help to answer in cancer cell biology are plentiful. To build on from the examples given in this paper, the methodology can be used to study interactions between cancer cells and a wide range of cancer-relevant cell types such as cancer-activated fibroblasts, macrophages, and other immune cells when co-cultured. Since a substantial proportion of cancer cells in tumors are in close proximity to other cell types that have been shown to affect tumor progression, using these co-cultures is more representative of the situation in a patient compared to studying cancer cells on their own. In addition to just giving the final cell number, the presented approach can dissect which cell types affect the growth of others and to what extent in complex heterogeneous populations. This could be relevant in a drug discovery setting to determine if a drug affects cancer cell growth due to internal effects (on other cancer cells) or by interfering with the interaction between the cancer cells and other cell types. Finding drugs with different targets and mechanisms of action are particularly sought after as they provide a wider target profile, increasing the chance of patients responding as well as reducing the risk of tumors becoming resistant. The impact of different genes and associated pathways in different cell types in relation to inter-cellular interactions can also be studied by genetically modifying the cell type(s) in question before mixing the cells together. This could be beneficial to identify new potential drug targets. Our approach is also applicable in other kinds of studies where local spatial cell-cell interactions are believed to affect cell growth such as studies of neurodegenerative diseases \cite{garden2012intercellular} and wound healing/tissue re-generation \cite{leoni2015wound}. In addition to evaluating cell growth, our approach can also be used to study transitions between cellular phenotypes upon interaction with other cell types, provided that the different phenotypes studied can be distinguished from one another based on the image data.
Finally, it is worth noting that issues may arise when cells become too confluent/dense, this may lead to segmentation problems of the imaging system. If they become completely confluent, they are likely to progressively stop growing. If one wants to measure for longer period of time, experiments can be performed in larger wells/plates or with smaller starting cell numbers.

Our methods offer several practical advantages to researchers interested in
analysing  multivariate count data on heterogeneous cell populations. First, the conditional Poisson model
does not require tracking  individual cells across time, a process that is often difficult to automate 
due to cell movement, morphology changes at subsequent time points, and additional complications related to storage of large data files. 
Second, we are able to quantify local spatio-temporal interactions between different cell populations from a very simple 
experimental set-up where the different cell populations are grown together in a single experimental condition (co-culture). An alternative, solely experimentally-based strategy would require monitoring the different
cell  types alone and together at different cell densities (number of cells per
condition) in order to make inferences in terms of potential interactions.
However, such an approach would give no possibility of evaluating the spatial relations
in the co-culture conditions and would still restrict the number of 
simultaneously tested cell types to two. 

In the future, we foresee several useful extensions of the current methodology, possibly enabling the treatment of more complex experimental settings. 
First, complex experiments involving a large number of cell populations, $n_{\Ccal}$, would imply an over-parametrized model.
Clearly, this large number of parameters would be detrimental to both statistical accuracy and reliable optimization of the likelihood objective function  $\ell_n(\theta)$ (\ref{cl}). To address these issues, we plan to explore a penalized
likelihood of form  $\ell_n(\theta) - \text{pen}_{\lambda}(\theta)$, where  $\text{pen}(\theta)$ is a nonnegative sparsity-inducing penalty function. 
For example, in a different likelihood setting, Bardic et al. \cite{bradic2011penalized} consider the $L_1$-type penalty 
 $\text{pen}(\theta) = \lambda \sum | \theta  |,$ $\lambda > 0$.
 Second, for certain experiments, it would be desirable to modify 
 the statistics in (\ref{intensity2}) to include additional information on cell growth such as the distance between 
heterogeneous cells, and covariates describing cell morphology. 
 Besides, it would be useful to consider tiling the microscope image into a hexagonal lattice, which is a more natural choice in real application, since the distance between neighbouring tiles would be more even than that of a regular lattice.  

\section*{Acknowledgements}

The authors wish to acknowledge support from the Australian National Health and Medical Research Council grants 1049561, 1064987 and 1069024 to Fr\'{e}d\'{e}ric 
Hollande. Christina M{\o}lck is supported by the Danish Cancer Society.

\bibliographystyle{abbrvnat}
\bibliography{biblio}

\end{document}